\documentclass[cmex10,reqno,letter]{amsart}
\usepackage{amssymb}  
\usepackage{graphicx}

\newcommand{\re}{{\mathbb{R}}}
\newcommand{\comp}{{\mathbb{C}}}
\newcommand{\dd}{{\mathrm{d}}}
\newcommand{\Z}{{\mathbb{Z}}}
\newcommand{\V}{V}
\newcommand{\Sr}{\mathcal{S}}
\newcommand{\sgn}{\mathrm{sgn}}
\newcommand{\ipr}[2]{\langle #1, #2 \rangle}
\newcommand{\vc}[1]{{\boldsymbol{#1}}}
\newcommand{\elll}{\ell^1\text{-}\ell^2}
\newcommand{\reg}[1]{{\bf{\Omega}}(#1)}
\newcommand{\norm}[1]{\lVert#1\rVert}
\newcommand{\jj}{{\mathrm{j}}}

\newcommand{\ltwoopt}{\vc{\theta}_{2}^\star}
\newcommand{\loneopt}{\vc{\theta}_{1}^\star}
\newcommand{\card}[1]{\mathrm{card}(#1)}
\newcommand{\plant}{\begin{cases}\dot{\vc{x}}=A\vc{x}+\vc{b}u\\y=\vc{c}^\top\vc{x}\end{cases}}
\newcommand{\red}{}

\newtheorem{prob}{Problem}
\newtheorem{thm}{Theorem}

\newtheorem{lem}{Lemma}
\newtheorem{defn}{Definition}

\newtheorem{ex}{Example}
\newtheorem{rem}{Remark}

\begin{document}

\title[Compressive Sampling for Remote Control Systems]{Compressive Sampling for Remote Control Systems%
\footnote{IEICE Trans.~on Fundamentals, Vol.~E95-A, No.~4, pp.~713--722 (2012)\hfill}
}
\author[M. Nagahara]{Masaaki Nagahara}
\author[T. Matsuda]{Takahiro Matsuda}
\author[K. Hayashi]{Kazunori Hayashi}
\address{M.~Nagahara and K. Hayashi are with
	Graduate School of Informatics, Kyoto University.
	T. Matuda is with Graduate School of Engineering, Osaka University.
}
\address{The corresponding author is Masaaki Nagahara.
	Mailing address: Graduate School of Informatics, Kyoto University, Yoshida Honmachi, Sakyo-ku, Kyoto 606-8501, Japan.
	}
\email{nagahara@ieee.org}


\maketitle
\begin{abstract}
In remote control, 
efficient compression or representation
of control signals is essential to send them through rate-limited channels.
For this purpose, we propose an approach of sparse control signal representation
using the compressive sampling technique.
The problem of obtaining sparse representation is formulated by
cardinality-constrained $\ell^2$ optimization of the control performance, which is reducible to
$\elll$ optimization.
The low rate random sampling employed in the proposed method based on the compressive sampling,
in addition to the fact that the $\elll$  optimization can be effectively solved by a fast iteration method, 
enables us to generate the sparse control signal with reduced computational complexity,
which is preferable in remote control systems where computation delays seriously degrade the performance.
{\red We give a theoretical result for control performance analysis based on the notion of 
restricted isometry property (RIP).}
An example is shown to illustrate the effectiveness of the proposed approach via numerical experiments.
\end{abstract}

\section{Introduction}
Remote control systems are
those in which the controlled objects are located away from
the control signal generators.
They are widely used at the present day,
from video games \cite{Lee08} to spacecraft \cite{Kub-etal06},
see \cite{Say} for other examples.
In remote control systems, control signals are to be transmitted through
rate-limited channels such as
wireless channels \cite{WinHol00} or the Internet
\cite{LuoChe00}.
In such systems, efficient signal compression or representation is essential to send
control signals through communication channels.
For this purpose, we propose 
an approach of sparse control signal representation using
 the {\em compressive sampling} technique \cite{Can06,CanWak08,Don06}
for remote control systems.

Compressive sampling, also known as compressed sensing,
is a technique for acquiring and reconstructing signals in the {\em sparse-land}
\cite{BruDonEla09}.
Signal acquisition and reconstruction is one of the fundamental issues
in signal processing.
In many applications, signals are analog (or continuous-time)
before they are acquired and converted to digital (or discrete-time) signals.
The problem is how to acquire and convert analog signals to digital ones
without much information distortion, such as aliasing.
A well-known and widely-used solution to this problem is
{\em Shannon's sampling theorem} \cite{Sha49,Uns00}.
This theorem gives an acquisition and reconstruction method for perfect reconstruction;
if the sampling rate is faster than twice the Nyquist rate, the maximum frequency contained in the original analog signal,
then the original signal can be perfectly reconstructed via sinc series.
On the other hand, 
in the sparse-land, signals are {\em sparse} or 
{\em compressible}
under a certain signal representation (e.g., Fourier or wavelet).
This sparsity assumption on signals is known to be valid for many real signals,
e.g., see examples in \cite{StaMurFad}.
Compressive sampling is based on this fact, by which
one can reconstruct the original signal with very high fidelity
from far fewer samples than what the conventional sampling theorem requires.
Hence signal acquisition and compression can be performed in much more efficient manner,
than the conventional scheme such as the image compression JPEG \cite{Wal91},
where one acquires the full signal, then transforms it into the frequency domain,
and finally discards most of them to obtain a compressed signal.

The purpose of this paper is to propose to use compressive sampling
for remote control systems.
Our contributions in this paper are as follows:
\begin{itemize}
\item We propose a new feed-forward-based remote control system
with compressive sampling.
\item The proposed system can efficiently compress the control signals
with sparse representation.
\item The design problem is formulated by $\elll$ optimization which can be
efficiently solved.
\end{itemize}

The theory of compressive sampling has been applied to not only signal processing but also
statistics \cite{CanTao07}, information theory \cite{SarBarBar06},
machine learning \cite{CalJafSch09}, and so on.
For theory and application of compressive sampling, 
see books \cite{Mal,Ela,StaMurFad}.
However, to the best of our knowledge, so far only a few studies have applied compressive sampling
to {\em control}:
\cite{BhaBas11} proposes to use compressive sensing in feedback control systems
for perfect state reconstruction and
\cite{NagQue11} proposes sparse representation of transmitted control packets
for feedback control.
For remote control systems, \cite{NagQueOstMatHay11} also proposes to use $\elll$ optimization
as in this paper, but the compressive sampling technique (Fourier expansion and random sampling)
is not used.
As we mentioned above,
it is desirable that signals in remote control systems are effectively acquired and compactly compressed.
Therefore, we propose to adopt compressive sampling technique to remote control systems.

Compressive sampling in this paper can be considered as a kind of lossy compression.
In many lossy data compression problems, the objective 
is to find efficient approximate representations 
of the original data~\cite{Kon00},
and the distortion is measured by
the signal reconstruction error.
On the other hand, in this paper, we consider a different aspect of the
distortion, that is, we measure the efficiency of the lossy
compression with {\em control performance}.
In other words, our method aims at optimizing the control performance, e.g., minimizing 
the tracking error, while
usual compressive sampling minimizes the $\ell^2$ norm of the reconstruction error,
with a sparsity constraint.
This is a natural notion in control; we do not care about how small the compression
error of the control signal is
but how good the control performance is.
Thus we call the proposed approach {\em control-oriented compressive sampling}.

In remote control systems, control delays due to heavy computation seriously degrade the performance.
In compressive sampling, signal acquisition is realized by a random non-uniform sampler
\cite{CanWak08}
or a random demodulator \cite{Tro+10},
which takes almost no computational time.
In contrast, obtaining sparse representation of a signal is achieved by solving 
$\elll$ optimization \cite{ZibEla10}, also known as LASSO \cite{Tib96} or
basis pursuit de-noising \cite{CheDonSau98}.
The solution to the $\elll$ optimization cannot be represented in an analytical form
as in $\ell^2$ optimization, and hence
we resort to iteration method to achieve the optimal solution.
There have been recently a number of researches on this type of optimization,
and there are several efficient algorithms for the solution \cite{DauDefMol04,BecTeb09,ZibEla10}.
Moreover, the low rate random sampling leads to reduced computational complexity of optimization.
That is,
we can use such computationally efficient algorithms with the low rate sampling in remote control to reduce control delays.

The paper is organized as follows:
In Section \ref{sec:problem}, we define our control problem.
In Section \ref{sec:conventional}, we formulate and solve the problem
via conventional sampling theorem.
In Section \ref{sec:CS}, 
we propose a new control method based on compressive sampling.
{\red
Section \ref{sec:analysis}
gives a theoretical result for performance analysis of the proposed control.
}
We show a numerical example in Section \ref{sec:examples} to illustrate the effectiveness of the proposed method.
Finally, we make a conclusion in Section \ref{sec:conclusion}.

\subsection*{Notation}
In this paper, we use the following notation. 
$\Z$, $\re$ and $\comp$ 
denote the sets of integral,
real and complex numbers, respectively.
$\re^n$ and $\re^{m\times n}$ ($\comp^n$ and $\comp^{m\times n}$)
denote the sets of $n$-dimensional real (complex)
vectors and $m\times n$ matrices, respectively.
We use $\jj$ for the imaginary unit in $\comp$.
For a complex number $z\in\comp$,
$\bar{z}$ and ${\mathrm{Re}}~z$ represent 
the conjugate and the real part of $z$, respectively.
For a matrix (a vector) $M$, $M^\top$ and $M^\ast$ represent the transpose
and the Hermitian conjugate of $M$, respectively.
For a vector $\vc{v}=[v_1,\ldots,v_n]^\top \in\comp^n$,
we define $\ell^{0}$ ``norm'' $\|\vc{v}\|_0$  of $\vc{v}$ as
the number of the nonzero elements in $\vc{v}$,
and also define $\ell^{1}$ , $\ell^{2}$, and $\ell^{\infty}$ norms as
\[
 \|\vc{v}\|_1:=\sum_{i=1}^n |v_i|,~~\|\vc{v}\|_2:=\sqrt{\vc{v}^\ast\vc{v}},~~\|\vc{v}\|_\infty := \max_{i=1,\ldots,n} |v_i|,
\]
respectively.
For a finite set $I=\{I_1,\ldots,I_K\}\subset\Z$, we define $|I|:=K$.
We denote
by  $L^2[0,T]$ the Lebesgue space consisting of all square
integrable functions on $[0,T]\subset\re$, endowed with the
inner product
\[
 \ipr{x}{y} := \int_0^T x(t)\overline{y(t)}~\dd t, \quad x, y \in L^2[0,T],
\]
and the $L^2$ norm $\|x\|:=\sqrt{\ipr{x}{x}}$.
\section{Control Problem}
\label{sec:problem}
In this paper, we consider a control problem of a linear system $P$
on a finite time interval (or horizon) $[0,T]$, $T>0$,
given by
\begin{equation}
 P:\left\{ \begin{split}
	    \dot{\vc{x}}(t) &= A\vc{x}(t) + \vc{b}u(t),\\ 
	   y(t) &= \vc{c}^\top \vc{x}(t), \quad \vc{x}(0)=\vc{x}_0\in\re^\nu, \quad t\in[0,T],
 \end{split}\right.
 \label{eq:plant}
\end{equation}
where $A\in\re^{\nu\times \nu}$, $\vc{b},\vc{c} \in\re^{\nu\times 1}$.
In this equation, $\vc{x}(t)\in\re^{\nu}$ is the state, 
$u(t)\in\re$ is the input, and $y(t)\in\re$ is the output of the system $P$.
The initial state $\vc{x}_0\in\re^\nu$ is assumed to be given.
We also assume that the system is stable, that is, 
the eigenvalues of $A$ are in $\comp_-=\{\lambda\in\comp:{\mathrm{Re}}~\lambda < 0\}$.
Then the system $P$ can be considered as a bounded operator in $L^2[0,T]$ for any $T>0$.
We use the notation $y=Pu$ for representing the input/output relation of the
linear system $P$.
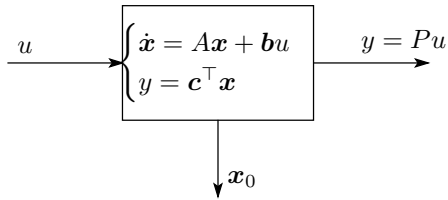
\begin{figure}[tb]
\centering{
\unitlength 0.1in
\begin{picture}( 22.0000, 10.0000)(  2.0000,-10.0000)
%
\special{pn 8}%
\special{pa 800 0}%
\special{pa 1800 0}%
\special{pa 1800 600}%
\special{pa 800 600}%
\special{pa 800 0}%
\special{fp}%
%
\special{pn 8}%
\special{pa 1800 300}%
\special{pa 2400 300}%
\special{fp}%
\special{sh 1}%
\special{pa 2400 300}%
\special{pa 2334 280}%
\special{pa 2348 300}%
\special{pa 2334 320}%
\special{pa 2400 300}%
\special{fp}%
%
\special{pn 8}%
\special{pa 200 300}%
\special{pa 800 300}%
\special{fp}%
\special{sh 1}%
\special{pa 800 300}%
\special{pa 734 280}%
\special{pa 748 300}%
\special{pa 734 320}%
\special{pa 800 300}%
\special{fp}%
\put(2.5000,-2.5000){\makebox(0,0)[lb]{$u$}}%
\put(20.5000,-2.5000){\makebox(0,0)[lb]{$y=Pu$}}%
%
\special{pn 8}%
\special{pa 1300 600}%
\special{pa 1300 1000}%
\special{fp}%
\special{sh 1}%
\special{pa 1300 1000}%
\special{pa 1320 934}%
\special{pa 1300 948}%
\special{pa 1280 934}%
\special{pa 1300 1000}%
\special{fp}%
\put(13.5000,-9.5000){\makebox(0,0)[lb]{$\vc{x}_0$}}%
\put(13.0000,-3.0000){\makebox(0,0){$\plant$}}%
\end{picture}%

}
\caption{Linear system $P$ to be controlled. The control signal $u$ is transmitted through a communication channel.
The initial state $\vc{x}_0$ is assumed to be measured.}
\label{fig:plant}
\end{figure}
Fig.~\ref{fig:plant} shows the block diagram of the system $P$ with the input $u$,
the output $y=Pu$, and the initial state $\vc{x}_0$.

In order to show the significance of the proposed approach,
we consider {\em tracking problem} in this paper as an example of the control problem.
In the tracking problem,
the controller attempts to reduce the tracking error between
a given reference $r$ and the output $y=Pu$ over $[0,T]$.
In other words, we design a control signal $\{u(t)\}_{t\in[0,T]}$
for a reference signal $\{r(t)\}_{t\in[0,T]}$
such that $r\approx Pu$ over $[0,T]$.
More precisely, the control object is described as follows:
Find a control signal $\{u(t)\}_{t\in[0,T]}$ such that
\begin{enumerate}
\item the tracking error $E(u):=\norm{Pu-r}^2$ is small,
\item the ``size'' $\reg{u}$ of the control signal $u$ is not too large,
\item and the maximum frequency contained in $u$ is bounded by a fixed frequency.
\end{enumerate}

The first objective is for tracking performance;
if $E(u)$ is smaller, the performance is said to be better.
Theoretically, $E(u)$ can be made arbitrarily small
if the size $\reg{u}$ is not restricted.
\begin{ex}
\label{ex:step}
Let $\hat{P}(s)$ denote the Laplace transform of the impulse response of 
the linear system $P$.
Suppose $\hat{P}(s)$ is given by
\[
 \hat{P}(s)=\frac{s-\alpha}{s+\alpha},
\]
where $\alpha>0$,
and the reference $r$ is given in the Laplace transform by
$\hat{r}(s) = 1/s$.
Then if we choose $u$ with its Laplace transform
\[
 \hat{u}(s) = \hat{P}(s)^{-1}\hat{r}(s) = \frac{s+\alpha}{s(s-\alpha)},
\]
then the performance in terms of the tracking error will be perfect, that is, $E(u) = 0$ over $[0,T]$.
However, the inverse Laplace transform of $\hat{u}(s)$ is given by
$u(t)=2\exp(\alpha t)-1$, $t\in [0,\infty)$,
and hence $u(t)$ has the property 
$\lim_{t\rightarrow\infty}u(t)=\infty$ since $\alpha>0$.
That is, if $T$ becomes large, then $|u(T)|$ increases exponentially.
\hfill$\Box$
\end{ex}

This example is not a special case;
we can generally say that a small tracking error
leads to a large control signal if $\hat{P}(s)$ has an unstable zero,
that is, there exists $z\in\comp_+=\{z\in\comp:{\mathrm{Re}}~z \geq 0\}$
such that $\hat{P}(z)=0$.
For example, suppose that the size of $u$ is measured
by $\reg{u}=\norm{u}^2$,
the energy of the control signal $u$.
As mentioned above, a smaller tracking error
$E(u)$ leads to a larger energy $\reg{u}=\norm{u}^2$.
It follows that we have to transmit the information of a signal
with a very large energy through a communication channel.
In many cases, a larger energy results in a larger amplitude of a signal,
and hence the variance becomes larger if the mean of $u(t)$ is $0$.
This implies that the entropy of the signal increases and so does the amount of information.
Moreover, large $\reg{u}$ leads to high sensitivity to
noise in measurement of the initial value $\vc{x}_0$ or uncertainty in the model parameters
$A$, $\vc{b}$, and $\vc{c}$.
We therefore add a constraint on $\reg{u}$ as the second control objective.
The size $\reg{u}$ is not restricted to the energy;
one can take another function as will be defined in Section \ref{sec:CS}.

The third objective is also needed in real control systems.
The control signal $u$ is applied to the controlled object
through an actuator (e.g., a motor),
which cannot act at a speed faster than a fixed frequency.
To describe this constraint mathematically,
we define a subspace of $L^2[0,T]$ by
\[
 \V_M := \mathrm{span}\{\psi_m:m=-M,\ldots,M\} \subset L^2[0,T],
\]
where $M$ is a given positive integer and
\[
 \psi_m := \frac{1}{\sqrt{T}}\exp({\jj\omega_m t}),\quad \omega_m := \frac{2\pi m}{T}.
\]
$\V_M$ is the set of $T$-periodic band-limited signals up to the frequency $\omega_M=2\pi M/T$ [rad/sec].
We restrict the control signal $u$ and the reference $r$ to this subspace.

The control problem considered in this paper is summarized as follows:
\begin{prob}[Tracking control problem]
Given a reference signal $r\in\V_M$,
find a control signal $u\in\V_M$ which minimizes
\begin{equation}
 J(u)=\norm{Pu-r}^2 + \mu \reg{u}
 \label{eq:2norm-const}
\end{equation}
where $\mu$ is a positive parameter which controls the tradeoff between $\norm{Pu-r}^2$ and $\reg{u}$.
\end{prob}

{\red
If the regularization term $\reg{u}$ in (\ref{eq:2norm-const}) is
defined as $\reg{u}\equiv 0$,
the optimization problem becomes the least-square optimization.
The solution is ideal in the sense that this gives the least squared error,
as the controller given in Example \ref{ex:step}.
As mentioned above, this ideal control may have very large energy or amplitude,
and the {\em energy-saving} constraint $\reg{u}=\norm{u}^2$
is conventionally used (see Section \ref{sec:conventional}).
On the other hand, we propose to use another constraint,
{\em sparsity-promoting} constraint, $\reg{u}=\card{u}$,
where $\card{u}$ is the cardinality (or sparsity)
of the signal $u$, which is mathematically defined in Section
\ref{sec:CS}.
We sum up these regularization terms in Table ~\ref{tbl:regularization}.
\begin{table}[t]
\caption{\red Regularization term}
\label{tbl:regularization}
\begin{center}
\begin{tabular}{|c|c|}
\hline
\red $\reg{u}$ & \red Purpose\\\hline
\red $0$  & \red Least squared error (ideal)\\
\red $\norm{u}^2$ & \red Energy-saving (conventional) \\
\red $\card{u}$ & \red Sparsity-promoting (proposed)\\
\hline
\end{tabular}
\end{center}
\end{table}
}

\section{Conventional Approach via Sampling Theorem}
\label{sec:conventional}
A conventional solution to the problem is obtained by the sampling theorem \cite{Sha49,Uns00}.
First, since the signals $r$ and $u$ are band-limited up to the frequency $\omega_M$,
we may safely sample the signals $r$ and $y=Pu$ at a rate faster than the Nyquist rate $2\omega_M$,
based on the sampling theorem.
Then, we define the {\em sampled} error functional
\[
  E_{\dd}(u) = h\sum_{n=1}^N |y(t_n)-r(t_n)|^2
  = h\sum_{n=1}^N |(Pu)(t_n)-r(t_n)|^2,
\]
where $N:=2M+1$ is the number of sampled data,
$h:=T/(N-1)$ the sampling period, and
$t_n:=(n-1)h$  the $n$-th sampling instant.
Then we assume $u\in \V_M$, that is, $u$ is represented by
\begin{equation}
 u = \sum_{m=-M}^M \theta_m \psi_m,
\label{eq:control}
\end{equation}
where $\theta_m\in\comp$, $m=-M,\ldots,M$.
{\red
The following lemma gives the expression of the output $y$
in terms of the coefficients $\theta_m$.
}
\begin{lem}
\label{lem:output}
{\red
For the control $u$ given in (\ref{eq:control}), 
the output $y$ of the plant $P$ defined in (\ref{eq:plant})
is given by
\begin{equation}
 y(\tau) = \vc{c}^\top \exp(\tau A)\vc{x}_0 + \sum_{m=-M}^M \theta_m \langle \kappa(\tau,\cdot), \psi_m \rangle,~\tau\in[0,T],
 \label{eq:yt}
\end{equation}
where $\kappa(\tau,t)$ is defined by
\[
 \kappa(\tau,t) := \begin{cases}\vc{c}^\top \exp\left[{(\tau-t)A}\right]\vc{b},
	       \quad &\text{if } 0\leq t < \tau \leq T,\\
	       0,\quad &\text{otherwise.}
	      \end{cases}
\]
}
\end{lem}
{\red
{\bf Proof:} The proof is given in  \ref{sec:app:output}.
\hfill $\Box$

This lemma gives
the sampled output $y(t_n)$,
$n=1,2,\ldots,N$, by
\begin{equation}
 y(t_n) = \vc{c}^\top \exp(t_nA)\vc{x}_0 + \sum_{m=-M}^M \theta_m \langle \phi_n, \psi_m \rangle,
 \label{eq:ytn}
\end{equation}
where $\phi_n=\kappa(t_n,\cdot)$, $n=1,2,\ldots,N$.
}
Note that the function $\phi_n$ is known as the control theoretic spline \cite{SunEgeMar00}.
By this, the sampled error functional $E_\dd(u)$ is described in terms of $\vc{\theta}:=[\theta_{-M},\ldots,\theta_{M}]^\top\in\comp^N$:
\[
 E_\dd\left(\sum_{m=-M}^M \theta_m \psi_m\right) = h\left\|G\vc{\theta}-H\vc{x}_0-\vc{r}\right\|_2^2,
\]
where 
\begin{equation}
 G := \begin{bmatrix}
	\ipr{\phi_1}{\psi_{-M}}&\ldots&\ipr{\phi_1}{\psi_M}\\
	\ipr{\phi_2}{\psi_{-M}}&\ldots&\ipr{\phi_2}{\psi_M}\\
	\vdots&\ddots&\vdots\\
	\ipr{\phi_N}{\psi_{-M}}&\ldots&\ipr{\phi_N}{\psi_M}
  \end{bmatrix} \in \comp^{N\times N},
 \label{eq:grammian}
\end{equation}
\[
 \vc{r} := \begin{bmatrix}r(t_1)\\r(t_2)\\\vdots\\r(t_N)\end{bmatrix}\in\re^N,\quad
 H :=  	\begin{bmatrix}\vc{c}^\top \exp\left({t_1A}\right)\\\vc{c}^\top \exp\left({t_2A}\right)\\\vdots\\\vc{c}^\top \exp\left({t_NA}\right)\end{bmatrix}
  \in \re^{N\times \nu}.
\]
The regularization term $\reg{u}$ is in this case naturally taken by
\[
 \reg{u}=\norm{u}^2=\|\vc{\theta}\|_2^2,
\]
where the second equality is due to Parseval's identity \cite{Hsu}

Finally, the problem is described as follows:
\begin{prob}[$\ell^2$ optimization]
Find a vector $\vc{\theta}\in\comp^N$
which minimizes the following cost functional:
\begin{equation}
 J_2(\vc{\theta}) := \|G\vc{\theta}-\vc{\beta}\|_2^2+\mu_2\|\vc{\theta}\|_2^2,
 \label{eq:J2}
\end{equation}
where $\vc{\beta}:=\vc{r}-H\vc{x}_0$ and $\mu_2:=\mu/h$.
\end{prob}
The solution of the above problem is given by \cite{SchSmo}
\begin{equation}
 \ltwoopt = (\mu_2 I + G^\top G)^{-1}G^{\top}\vc{\beta}.
 \label{eq:ltwoopt}
\end{equation}
Thus, in conventional approach,
all the elements of $\ltwoopt$ (or $N$ samples of the 
corresponding control signal $u$) will be sent through a  rate-limited communication channel.
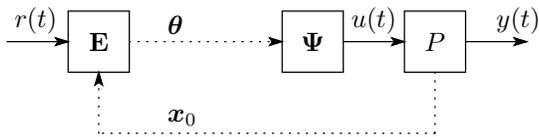
\begin{figure}[tb]
\centering{
\unitlength 0.1in
\begin{picture}( 28.2500,  6.9000)(  2.9500, -8.3600)
%
\special{pn 8}%
\special{pa 400 356}%
\special{pa 720 356}%
\special{fp}%
\special{sh 1}%
\special{pa 720 356}%
\special{pa 654 336}%
\special{pa 668 356}%
\special{pa 654 376}%
\special{pa 720 356}%
\special{fp}%
%
\special{pn 8}%
\special{pa 720 196}%
\special{pa 1040 196}%
\special{pa 1040 516}%
\special{pa 720 516}%
\special{pa 720 196}%
\special{fp}%
%
\special{pn 8}%
\special{pa 1040 356}%
\special{pa 1840 356}%
\special{dt 0.045}%
\special{sh 1}%
\special{pa 1840 356}%
\special{pa 1774 336}%
\special{pa 1788 356}%
\special{pa 1774 376}%
\special{pa 1840 356}%
\special{fp}%
%
\special{pn 8}%
\special{pa 1840 196}%
\special{pa 2160 196}%
\special{pa 2160 516}%
\special{pa 1840 516}%
\special{pa 1840 196}%
\special{fp}%
%
\special{pn 8}%
\special{pa 2160 356}%
\special{pa 2480 356}%
\special{fp}%
\special{sh 1}%
\special{pa 2480 356}%
\special{pa 2414 336}%
\special{pa 2428 356}%
\special{pa 2414 376}%
\special{pa 2480 356}%
\special{fp}%
%
\special{pn 8}%
\special{pa 2480 196}%
\special{pa 2800 196}%
\special{pa 2800 516}%
\special{pa 2480 516}%
\special{pa 2480 196}%
\special{fp}%
%
\special{pn 8}%
\special{pa 2800 356}%
\special{pa 3120 356}%
\special{fp}%
\special{sh 1}%
\special{pa 3120 356}%
\special{pa 3054 336}%
\special{pa 3068 356}%
\special{pa 3054 376}%
\special{pa 3120 356}%
\special{fp}%
%
\special{pn 8}%
\special{pa 2640 516}%
\special{pa 2640 836}%
\special{dt 0.045}%
\special{pa 2640 836}%
\special{pa 880 836}%
\special{dt 0.045}%
%
\special{pn 8}%
\special{pa 880 836}%
\special{pa 880 516}%
\special{dt 0.045}%
\special{sh 1}%
\special{pa 880 516}%
\special{pa 860 584}%
\special{pa 880 570}%
\special{pa 900 584}%
\special{pa 880 516}%
\special{fp}%
\put(4.4000,-3.1600){\makebox(0,0)[lb]{$r(t)$}}%
\put(12.4000,-3.1600){\makebox(0,0)[lb]{$\vc{\theta}$}}%
\put(22.0000,-3.1600){\makebox(0,0)[lb]{$u(t)$}}%
\put(29.6000,-3.1600){\makebox(0,0)[lb]{$y(t)$}}%
\put(12.4000,-7.9600){\makebox(0,0)[lb]{$\vc{x}_0$}}%
\put(8.8000,-3.5600){\makebox(0,0){${\mathbf E}$}}%
\put(20.0000,-3.5600){\makebox(0,0){${\mathbf \Psi}$}}%
\put(26.4000,-3.5600){\makebox(0,0){$P$}}%
\end{picture}%
}
\caption{
Remote control system}
\label{fig:rcs}
\end{figure}
Fig.~\ref{fig:rcs} shows the remote control system considered here.
In this figure, a continuous-time signal is drawn by a continuous arrow and
a transmitted vector by a dotted arrow.
The function ${\mathbf E}$ maps the reference $\{r(t)\}_{t\in[0,T]}$ and the initial state $\vc{x}_0$ of the system $P$
to the optimal vector $\vc{\theta}=\ltwoopt$ using (\ref{eq:ltwoopt}),
and the computed $\vc{\theta}$ is 
encoded and transmitted through the channel.
Then the signal $\vc{\theta}$ is received at ${\mathbf \Psi}$ which converts $\vc{\theta}$ to the control signal $\{u(t)\}_{t\in[0,T]}$
via the Fourier expansion as in (\ref{eq:control}).
Finally, the control signal $u$ is added to the system $P$.

\section{Proposed Approach via Compressive Sampling}
\label{sec:CS}
We here propose sparse representation of transmitted vector $\vc{\theta}$ in Fig.~\ref{fig:rcs}
for data compression via compressive sampling.
\subsection{Proposed formulation using sparse representation}
\label{subsec:proposed}
As we have seen in Section \ref{sec:problem}, there is a trade-off between the performance and
size of the control signal, and in the conventional approach, the balance is taken by
employing $\ell^{2}$ norm as the definition of the size.
In order to further reduce the size (or the amount of information) of the control signal $u$,
while keeping a certain degree of the distortion $\norm{Pu-r}^2$,
we impose a stronger but acceptable assumption on signals, that is, {\em sparsity}.

We first assume that the reference $r\in\V_M$ is sparse with respect to the basis $\{\psi_m\}$,
that is, a few of the Fourier coefficients of $r$ are nonzero while the others are zero.
This is represented by
\[
 r = \sum_{m\in I} r_m \psi_m, \quad I \subset \{-M,\ldots,M\},\quad |I|=S_{\!r},
\]
where $S_{\!r}\ll N=2M+1$.
The sparsity assumption on the reference signal is realistic in actual control systems.
For example, the step reference $\hat{r}(s)=1/s$ in Example \ref{ex:step},
or a sinusoidal reference with one frequency or a sum of several sinusoids,
which are typical reference signals,
are all sparse in the Fourier expansion.
{\red
In general, it is difficult to find a proper basis with which
reference signals are sparse.
However, we fix the Fourier basis and assume the reference signals
are sparse in the Fourier domain.
Under this assumption,
checking the sparsity $S_{\!r}$ of a given reference signal $r$
can be performed by the following steps:
\begin{enumerate}
\item sample the reference $r(t)$ with sampling frequency $2\omega_M$,
\item compute the Fourier coefficients via FFT from the sampled data,
\item truncate small coefficients,
\item count the number of the nonzero coefficients.
\end{enumerate}
If the number is small enough relative to the size $N=2M+1$,
we can say the signal is sparse.
We have assumed that the reference $r$ is in the signal subspace $V_M$,
that is the reference is $T$-periodic
and band-limited up to the frequency $\omega_M$,
the above procedure should work well.
}
Under the above assumption, we then consider the control signal $u$ 
defined in (\ref{eq:control}).
In general, the optimal control signal $u$ may not be sparse
even if the reference $r$ is sparse
(see Example \ref{ex:step}).
Nevertheless, we propose to assume the control signal $u$ to be sparse 
by {\em designing}  $u$ to be sparse.
The validity of the approach could be justified as follows:
\begin{enumerate}
\item The coefficient vector $\vc{\theta}=[\theta_{-M},\ldots,\theta_M]^\top$ 
of the control signal
is transmitted through a rate-limited communication channel
(see Fig.~\ref{fig:rcs}).
A sparse vector is then more desirable than the full vector $\ltwoopt$ in (\ref{eq:ltwoopt})
from a view point of data compression.
\item We will adopt the $\ell^1$ norm minimization for $\vc{\theta}$
as a sparsity-promoting criterion in Section \ref{sec:CS}.
Then a small $\ell^1$ norm of $\vc{\theta}$ leads to a small $L^1$ norm of $u$ since
\[
 \int_0^T |u(t)|\dd t \leq \sum_{m=-M}^{M} |\theta_m| \int_0^T |\psi_m(t)|\dd t = T\|\vc{\theta}\|_1.
\]
Thus, the size of $u$ measured by $L^1$ norm can be made small.
It follows that it can gain robustness against noise and model uncertainty.
\item If the control input $u$ is sparse, then the output $y=Pu$ is also sparse at steady state.
In fact, by the theory of linear systems \cite{Hsu},
the steady state response $y_{\text{ss}}$ of $P$ for the input $u$ given in (\ref{eq:control}) becomes
\[
 y_{\text{ss}} = \sum_{m=-M}^m \hat{P}(\jj \omega_m) \theta_m \psi_m,
\]
where $\hat{P}(s)$ is the Laplace transform of the impulse response of $P$.
Therefore, if $\{\theta_m\}$ is sparse, so is $\{\hat{P}(\jj \omega_m)\theta_m\}$.
This fact endorses the sparsity constraint on the control signal $u$ when the reference $r$ is sparse.
\end{enumerate}

Now we formulate our problem.
We denote by $\card{u}$ the number of the nonzero Fourier coefficients with respect to 
the basis $\{\psi_m\}$.
If $u$ is represented as in (\ref{eq:control}), then
$\card{u}=\|\vc{\theta}\|_0$.
For promoting sparsity of the control signal $u$, we set the regularization term $\reg{u}=\card{u}$.
In summary, our problem is formulated as follows:
\begin{prob}[Sparsity-promoting optimization]
\label{prob:sparse-control}
Given a reference signal $r\in\V_M$ with $\card{r}=S\ll N$,
find a control signal $u\in \V_M$ which minimizes
\[
 J_0(u) := \norm{Pu-r}^2 + \mu~\card{u}.
\]
\end{prob}

\subsection{Random sampling and $\elll$ optimization}
The control signal $u$ can be obtained by using the sampled error functional
$ J_0(u) $  with the Nyquist rate sampling as in Section \ref{sec:conventional}, and by solving the optimization problem.
However, based on the idea of compressive sampling \cite{Can06,CanWak08,Don06},
we can obtain the sparse control signal with much reduced computational complexity.
Specifically, we adopt low rate random sampling of signals instead of the uniform Nyquist rate sampling.

Let $U$ be a random ``decimation'' matrix of the form
\[
 U = \begin{bmatrix}\vc{e}_{i(1)}\\\vc{e}_{i(2)}\\\vdots\\\vc{e}_{i(K)}\end{bmatrix} \in \{0,1\}^{K\times N},
\]
where $i(1)<i(2)<\cdots<i(K)$ are the random variables of the uniform distribution on $\{1,2,\ldots,N\}$, and
\[
  \vc{e}_i:=[0,\ldots,0, \stackrel{i}{\stackrel{\vee}{1}},0,\ldots,0], \quad i=1,2,\ldots,N.
\]
\begin{figure}[tb]
\centering{
\unitlength 0.1in
\begin{picture}( 23.3500, 13.6500)(  2.6500,-15.6500)
%
\special{pn 8}%
\special{pa 400 1400}%
\special{pa 400 200}%
\special{fp}%
\special{sh 1}%
\special{pa 400 200}%
\special{pa 380 268}%
\special{pa 400 254}%
\special{pa 420 268}%
\special{pa 400 200}%
\special{fp}%
%
\special{pn 8}%
\special{pa 400 1400}%
\special{pa 2600 1400}%
\special{fp}%
\special{sh 1}%
\special{pa 2600 1400}%
\special{pa 2534 1380}%
\special{pa 2548 1400}%
\special{pa 2534 1420}%
\special{pa 2600 1400}%
\special{fp}%
%
\special{pn 8}%
\special{pa 400 800}%
\special{pa 430 778}%
\special{pa 458 758}%
\special{pa 488 736}%
\special{pa 518 716}%
\special{pa 546 698}%
\special{pa 576 678}%
\special{pa 604 662}%
\special{pa 632 646}%
\special{pa 662 634}%
\special{pa 690 622}%
\special{pa 718 612}%
\special{pa 746 606}%
\special{pa 774 602}%
\special{pa 802 600}%
\special{pa 828 602}%
\special{pa 856 608}%
\special{pa 882 614}%
\special{pa 908 624}%
\special{pa 936 638}%
\special{pa 962 652}%
\special{pa 988 668}%
\special{pa 1014 686}%
\special{pa 1040 706}%
\special{pa 1066 726}%
\special{pa 1092 748}%
\special{pa 1118 770}%
\special{pa 1144 794}%
\special{pa 1170 816}%
\special{pa 1196 840}%
\special{pa 1222 864}%
\special{pa 1250 886}%
\special{pa 1276 910}%
\special{pa 1304 932}%
\special{pa 1330 952}%
\special{pa 1358 972}%
\special{pa 1386 992}%
\special{pa 1414 1008}%
\special{pa 1444 1024}%
\special{pa 1472 1038}%
\special{pa 1502 1050}%
\special{pa 1530 1062}%
\special{pa 1560 1072}%
\special{pa 1592 1080}%
\special{pa 1622 1086}%
\special{pa 1652 1092}%
\special{pa 1684 1096}%
\special{pa 1714 1100}%
\special{pa 1746 1102}%
\special{pa 1778 1104}%
\special{pa 1810 1104}%
\special{pa 1842 1104}%
\special{pa 1874 1102}%
\special{pa 1906 1098}%
\special{pa 1940 1096}%
\special{pa 1972 1092}%
\special{pa 2006 1088}%
\special{pa 2038 1082}%
\special{pa 2072 1076}%
\special{pa 2106 1070}%
\special{pa 2138 1064}%
\special{pa 2172 1056}%
\special{pa 2206 1048}%
\special{pa 2240 1040}%
\special{pa 2274 1032}%
\special{pa 2308 1024}%
\special{pa 2340 1016}%
\special{pa 2374 1008}%
\special{pa 2400 1000}%
\special{sp}%
%
\special{pn 8}%
\special{pa 2400 1000}%
\special{pa 2400 1400}%
\special{dt 0.045}%
%
\special{pn 8}%
\special{sh 0.600}%
\special{ar 800 600 50 50  0.0000000 6.2831853}%
%
\special{pn 8}%
\special{sh 0.600}%
\special{ar 1200 850 50 50  0.0000000 6.2831853}%
%
\special{pn 8}%
\special{sh 0.600}%
\special{ar 1370 980 50 50  0.0000000 6.2831853}%
%
\special{pn 8}%
\special{sh 0.600}%
\special{ar 1970 1090 50 50  0.0000000 6.2831853}%
%
\special{pn 8}%
\special{pa 800 650}%
\special{pa 800 1400}%
\special{dt 0.045}%
%
\special{pn 8}%
\special{pa 1200 1400}%
\special{pa 1200 900}%
\special{dt 0.045}%
%
\special{pn 8}%
\special{pa 1370 1400}%
\special{pa 1370 1030}%
\special{dt 0.045}%
%
\special{pn 8}%
\special{pa 1970 1400}%
\special{pa 1970 1140}%
\special{dt 0.045}%
\put(25.5000,-13.4000){\makebox(0,0)[lb]{$t$}}%
\put(24.0000,-14.5000){\makebox(0,0)[lt]{$T$}}%
\put(19.7000,-14.5000){\makebox(0,0)[lt]{$t_{i(4)}$}}%
\put(13.7000,-14.5000){\makebox(0,0)[lt]{$t_{i(3)}$}}%
\put(11.6000,-14.5000){\makebox(0,0)[lt]{$t_{i(2)}$}}%
\put(8.0000,-14.5000){\makebox(0,0)[lt]{$t_{i(1)}$}}%
\put(4.0000,-14.5000){\makebox(0,0)[lt]{$0$}}%
\put(4.0000,-16.5000){\makebox(0,0){$~$}}%
\end{picture}%

}
\caption{Random sampling}
\label{fig:random_sampling}
\end{figure}
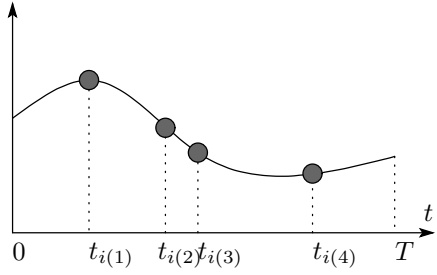
This is a model of low rate random sampling of a signal on $[0,T]$ as shown in Fig. \ref{fig:random_sampling},
where the sampling instants are given by
\[
 t_{i(k)}=i(k)\cdot h = i(k)\cdot \frac{T}{N-1},\quad k=1,2,\ldots,K<N.
\]
\begin{rem}
\label{rem:compression}
{\red
The choice of the number $K$ is a fundamental problem
in compressive sampling.
Suppose that the sparsity of the vector $\vc{\theta}$
is $\|\vc{\theta}\|_0=S_{\!\theta}$.
Then, for large $N$, one can choose $K$ as
$K \geq CS_{\!\theta}(\log N)^4$,
where $C$ is some constant \cite{RudVer08}.
It is believed that the bound may be reduced to
$CS_{\!\theta}(\log N)$, but there is no theoretical proof
\cite{Can06}.
}
\end{rem}

By using the matrix $U$, random sampling of $y(t)$ on $[0,T]$ is given by:
\[
 \vc{y} = UG\vc{\theta} + UH\vc{x}_0.
\]
Then the cost functional is given by 
\[
 J_0(\vc{\theta}) = \|\Phi\vc{\theta}-\vc{\alpha}\|_2^2 + \mu \|\vc{\theta}\|_0,
\]
where $\Phi:=UG$ and $\vc{\alpha}=U(\vc{r}-H\vc{x}_0)$.

It should be noted that, thanks to the low rate random sampling matrix $U$,
 the computational complexity of $J_0(\vc{\theta})$ is reduced compared as
 the case with the Nyquist rate uniform sampling.
However,  minimization of $J_0$ may still be hard to solve, because 
the optimization problem is a combinatorial one.
It is common to employ convex relaxation by replacing the
$\ell^0$ norm with the $\ell^1$ norm, thus we have
\begin{equation}
 J_1(\vc{\theta}) = \|\Phi\vc{\theta}-\vc{\alpha}\|_2^2 + \mu_1 \|\vc{\theta}\|_1.
 \label{eq:J1}
\end{equation}
The cost functional $J_1(\vc{\theta})$ in (\ref{eq:J1}) is convex in $\vc{\theta}$ and hence
the optimal value $\loneopt$ uniquely exists.
However, an analytical expression as in (\ref{eq:ltwoopt})
for this optimal vector
is unknown except when the matrix $\Phi$ is unitary.
To obtain the optimal vector $\loneopt$,
one can use an iteration method.
Recently, several fast algorithms to obtain the optimal $\elll$ solution
has been proposed, which is called {\it iterative shrinkage} \cite{BecTeb09,ZibEla10}.
In this paper, we use the algorithm called
FISTA (Fast Iterative Shrinkage-Thresholding Algorithm) \cite{BecTeb09}.
The algorithm converges to
the optimal solution minimizing the $\elll$ cost functional
(\ref{eq:J1}) for any initial guess of $\loneopt$
with a worst-case convergence rate $O(1/j^2)$ \cite{DauDefMol04,BecTeb09}.
The algorithm is very simple and fast;
it can be effectively implemented in digital devices,
which leads to a real-time computation of a sparse vector $\loneopt$.
For this algorithm, see \ref{sec:app:FISTA}.

In summary, the proposed remote control system with the structure in Fig.~\ref{fig:rcs}
employs the process ${\mathbf E}$ which maps $\{r(t)\}_{t\in[0,T]}$ and $\vc{x}_0$
to the $\elll$ optimal vector $\loneopt$ using FISTA.
Since the vector $\loneopt$ is sparse, one can encode the vector in a small size.
The transmitted signal $\loneopt$ is received at ${\mathbf \Psi}$ and
the control signal $\{u(t)\}_{t\in[0,T]}$ is obtained by (\ref{eq:control}).
Again since $\loneopt$ is sparse, this procedure can be efficiently done.

{\red
We have considered 3 cost functionals;
$J_2(\vc{\theta})$ in Section \ref{sec:conventional}, 
and $J_0(\vc{\theta})$ and $J_1(\vc{\theta})$ in this section.
We sum up these cost functionals in Table.~\ref{tbl:cost-functional}.
}
\begin{table}[t]
\caption{\red Cost functional}
\label{tbl:cost-functional}
\begin{center}
\begin{tabular}{|c|c|c|}
\hline
\red Cost & \red Purpose & \red Optimization\\\hline
\red $J_2(\vc{\theta})$  & \red Energy-saving (conventional)& \red Closed form solution\\
\red $J_0(\vc{\theta})$  & \red Sparsity-promoting (ideal)  &\red NP-hard\\
\red $J_1(\vc{\theta})$  & \red Sparsity-promoting (proposed) &\red Iteration\\
\hline
\end{tabular}
\end{center}
\end{table}

\section{\red Performance Analysis}
\label{sec:analysis}
{\red
In this section, we consider the performance analysis of
the proposed remote control systems.

Let $y^\star$ the ideal output of the plant $P$ under the control vector
$\vc{\theta}^\star$ which minimizes $\norm{Pu-r}^2$,
that is, $\reg{u}=0$ (See Table \ref{tbl:cost-functional}).
Let also $y^\star_1$ be the output with the proposed 
$\elll$-optimal vector 
$\vc{\theta}_1^\star$.
Clearly, the tracking performance by the $\elll$ optimal vector
$\vc{\theta}_1^\star$
is not better than that of the ideal $\vc{\theta}^\star$,
that is, 
\[
 \norm{y^\star-r}\leq \norm{y^\star_1-r}.
\]
The problem here is
to guarantee the boundedness of the tracking error
$\norm{y^\star_1-r}$ of the proposed $\elll$ control,
and to estimate the difference between the
two errors, $\norm{y^\star-r}$ and $\norm{y^\star_1-r}$,
when the errors are bounded.

Suppose that the ideal control vector $\vc{\theta}^\star$ is approximately $S$-sparse,
that is,
there exist a positive integer $S$ and a sufficiently small $\epsilon_1$ such that
\[
 \|\vc{\theta}^\star - \vc{\theta}^\star_{[S]}\|_1 \leq \epsilon_1,
\]
where $\vc{\theta}^\star_{[S]}$ is the vector $\vc{\theta}$ with all but
the largest $S$ components set to $0$.
Then, we introduce the notion of {\em restricted isometry property} (RIP)
\cite{Can06}.
\begin{defn}
For each integer $l=1,2,\ldots$, define the isometry constant
$\delta_l$ of a matrix $\Phi$ as the smallest number such that
\[
 (1-\delta_l)\|\vc{\theta}\|_2^2 \leq \|\Phi\vc{\theta}\|_2^2 \leq (1+\delta_l)\|\vc{\theta}\|_2^2
\]
holds for all vectors $\vc{\theta}$ such that $\|\vc{\theta}\|_0=l$.
\end{defn}
By using the notion of RIP, we have the following lemma:
\begin{lem}
\label{lem:analysis}
Assume that the isometry constant of the matrix $\Phi$ satisfies $\delta_{2S}<\sqrt{2}-1$.
Then, with sufficiently small $\mu_1>0$ in the cost functional $J_1(\vc{\theta})$
defined in (\ref{eq:J1}), we have the following estimate:
\begin{equation}
\|\vc{\theta}^\star_1-\vc{\theta}^\star\|_2 \leq C_1\frac{\epsilon_1}{\sqrt{S}} + C_2\epsilon_2,
\label{eq:theta-bound}
\end{equation}
where
\[
 \begin{split}
  C_1&:=2\cdot\frac{1+(\sqrt{2}-1)\delta_{2S}}{1-(\sqrt{2}+1)\delta_{2S}},~~
  C_2:=\frac{4\sqrt{1+\delta_{2S}}}{1-(\sqrt{2}+1)\delta_{2S}},\\
  \epsilon_2 &:= \|\Phi\vc{\theta}^\star_1 - \vc{\alpha}\|_2.
 \end{split}
\]
\end{lem}
{\bf Proof:}  The proof is given in  \ref{sec:app:analysis}.
\hfill $\Box$

By this lemma, we obtain the following bound
for tracking error by the $\elll$ optimal control.
\begin{thm}
\label{thm:analysis}
Assume $\delta_{2S}<\sqrt{2}-1$.
Then we have
\[
 \norm{y^\star_1-r} \leq \norm{y^\star-r} 
  + \left(C_0\frac{\epsilon_1}{\sqrt{S}} + C_1\epsilon_2\right)\eta,
\]
where
\[
 \eta := \sqrt{\sum_{m=-M}^M \int_0^T |\ipr{\kappa(\tau,\cdot)}{\psi_m}|^2 \dd \tau}.
\]
\end{thm}
{\bf Proof:}
By Lemma \ref{lem:output}, for $\tau \in [0,T]$, we have
\[
 y^\star_1(\tau) - y^\star_1(\tau) = \sum_{m=-M}^M \left(\theta_{1,m}^\star-\theta_m^\star\right)\ipr{\kappa(\tau,\cdot)}{\psi_m},
\]
where $\theta_{1,m}^\star$ and $\theta_m^\star$ are respectively the $m$-th components of $\vc{\theta}^\star_1$ and $\vc{\theta}^\star$.
Then, the Cauchy-Schwartz inequality \cite{You} gives
\[
 |y^\star_1(\tau)-y^\star(\tau)|^2 \leq \|\vc{\theta}_1^\star-\vc{\theta}^\star\|_2^2 \sum_{m=-M}^M |\ipr{\kappa(\tau,\cdot)}{\psi_m}|^2.
\]
It follows that
\[
 \begin{split}
  \norm{y^\star_1-y^\star}
  &= \sqrt{\int_0^T |y^\star_1(\tau)-y^\star(\tau)|^2\dd\tau}\\
  &\leq \sqrt{\int_0^T \|\vc{\theta}_1^\star-\vc{\theta}^\star\|_2^2 \sum_{m=-M}^M |\ipr{\kappa(\tau,\cdot)}{\psi_m}|^2\dd\tau}\\
  &= \|\vc{\theta}_1^\star-\vc{\theta}^\star\|_2\cdot\eta\\
  &\leq \left(C_0\frac{\epsilon_1}{\sqrt{S}}+C_1\epsilon_2\right)\eta.
 \end{split}
\]
The last inequality is due to Lemma \ref{lem:analysis}.
Finally, we have
\[
 \begin{split}
  \norm{y^\star_1-r} 
   &=\norm{y^\star_1-y^\star+y^\star-r}\\
   &\leq \norm{y^\star-r} + \norm{y^\star_1-y^\star}\\
   &\leq \norm{y^\star-r} + \left(C_0\frac{\epsilon_1}{\sqrt{S}}+C_1\epsilon_2\right)\eta.
 \end{split}
\]
\hfill $\Box$

By this theorem, we conclude that
the tracking error of the proposed $\elll$ optimal control is bounded
if $\norm{y^\star-r}$, the ideal control error, is bounded.
We also argue that the difference between the two performances, 
the ideal $\norm{y^\star-r}$
and the proposed $\norm{y^\star_1-r}$, is not so large if $\epsilon_1$ and $\epsilon_2$ 
are sufficiently small.
}

\section{Numerical Results}
\label{sec:examples}
We here give numerical examples to show the effectiveness of the proposed method.
The matrices of the system $P$ defined in (\ref{eq:plant}) to be controlled are given by
\[
 A = \begin{bmatrix}0&1\\-\alpha&-\alpha-1\end{bmatrix},~~
 \vc{b}=\begin{bmatrix}0\\1\end{bmatrix},~~
 \vc{c}=\begin{bmatrix}-\alpha\\1\end{bmatrix},
\]
with $\alpha=0.5$.
Note that the Laplace transform $\hat{P}(s)$ is 
\[
 \hat{P}(s) = \frac{s-\alpha}{(s+\alpha)(s+1)}.
\]
and this system has an unstable zero at $s=\alpha=0.5$ as mentioned in Example \ref{ex:step}.
We assume the initial state $\vc{x}_0=[0,0]^\top$.
The period $T$ is $2\pi$.
The number of basis $\{\psi_m\}$ is $N=2M+1=201$ ($M=100$).
The reference signal $r(t)$ is given by
\[
 r(t) = \sin(20t) + \cos(50t),
\]
and the sparsity (cardinality) of this reference is $S_{\!r}=4$.
For compressive sampling, we take $K=201/3=67$ random samples
among $N=201$ sampled data,
that is the compression ratio is $1/3$.

\begin{figure}[tb]
\centering{
\includegraphics[width=0.96\linewidth]{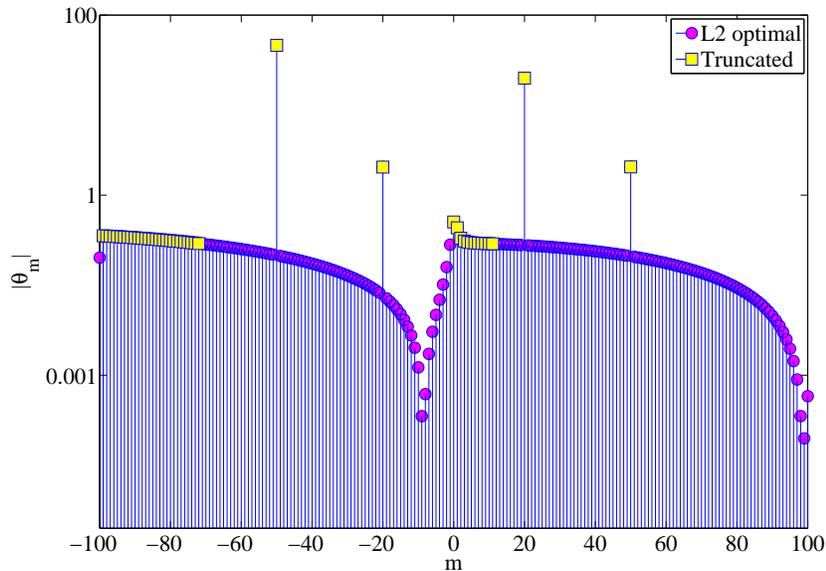}
}
\caption{
The absolute values of the elements of the Fourier coefficient vector $\ltwoopt$ in the $\ell^2$ optimal control signal $u\in\V_M$.
The squared markers show the $44$ largest coefficients which are used for a truncated vector.}
\label{fig:theta2}
\end{figure}

\begin{figure}[tb]
\centering{
\includegraphics[width=0.96\linewidth]{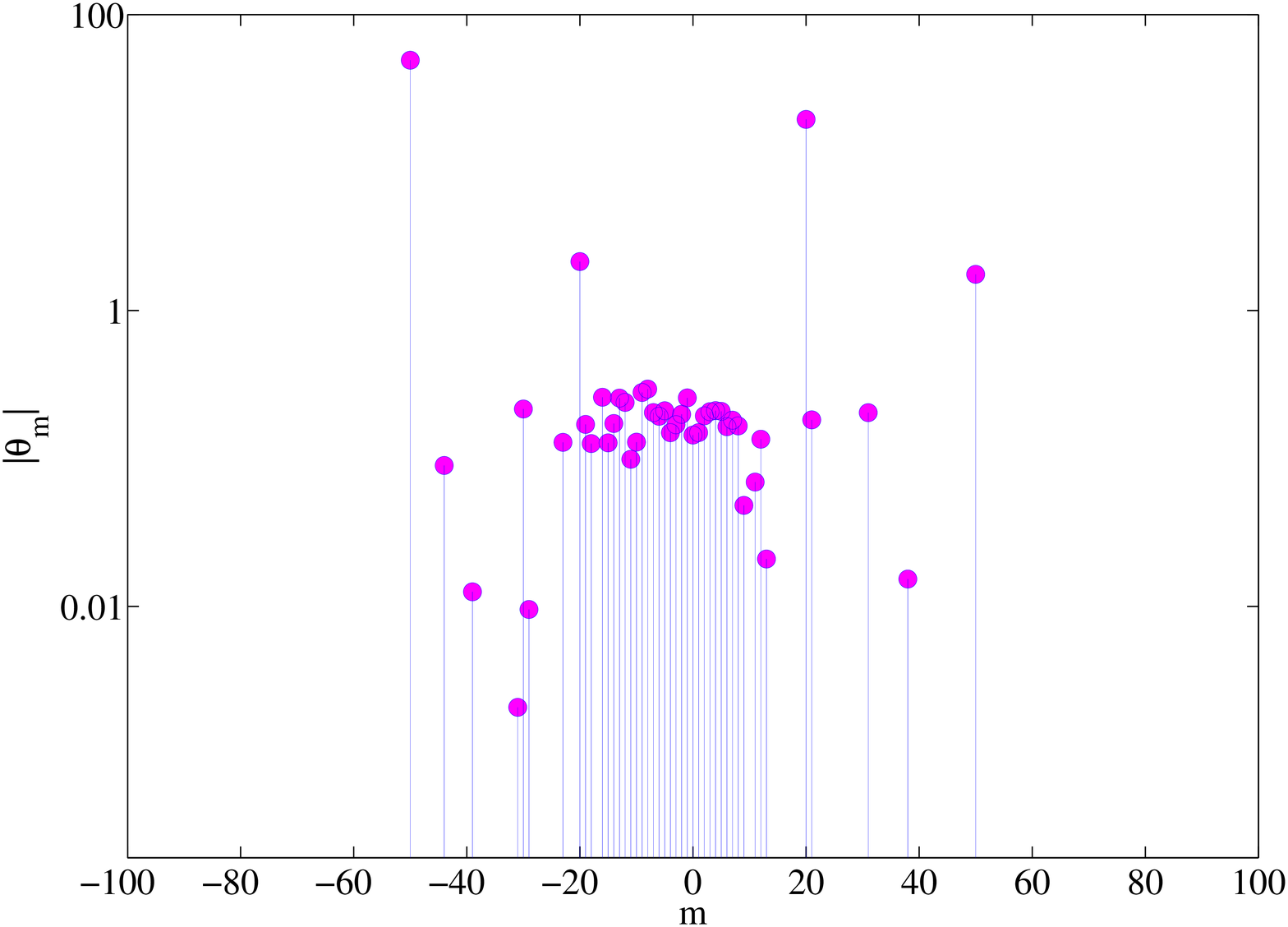}
}
\caption{
The absolute values of the elements of the Fourier coefficient vector $\loneopt$ in the $\elll$ optimal control signal $u\in\V_M$.
The $0$-valued elements are omitted. The sparsity is $\|\loneopt\|_0=44$.}
\label{fig:theta1}
\end{figure}

We compute the $\ell^2$ optimal Fourier coefficient vector $\ltwoopt$
minimizing (\ref{eq:J2}), given by (\ref{eq:ltwoopt}),
as a conventional design.
We also compute the $\elll$ optimal vector $\loneopt$ minimizing
(\ref{eq:J1}) as the proposed method.
The regularization parameters $\mu_1$ and $\mu_2$ respectively for 
$\elll$  and $\ell^2$ optimization are set to $\mu_1=\mu_2=10^{-4}$.
Fig.~\ref{fig:theta2} shows the elements of the vector $\ltwoopt$.
We can see that 4 elements are much larger than the other.
This vector however is not sparse, that is, $\|\ltwoopt\|_0=201$ (full).
On the other hand, Fig.~\ref{fig:theta1} shows the $\elll$ optimal $\loneopt$
which is very sparse. In fact, the sparsity is $\|\loneopt\|_0=44$,
about $21.9\%$ of the full vector $\ltwoopt$.

\begin{figure}[tb]
\centering{
\includegraphics[width=0.96\linewidth]{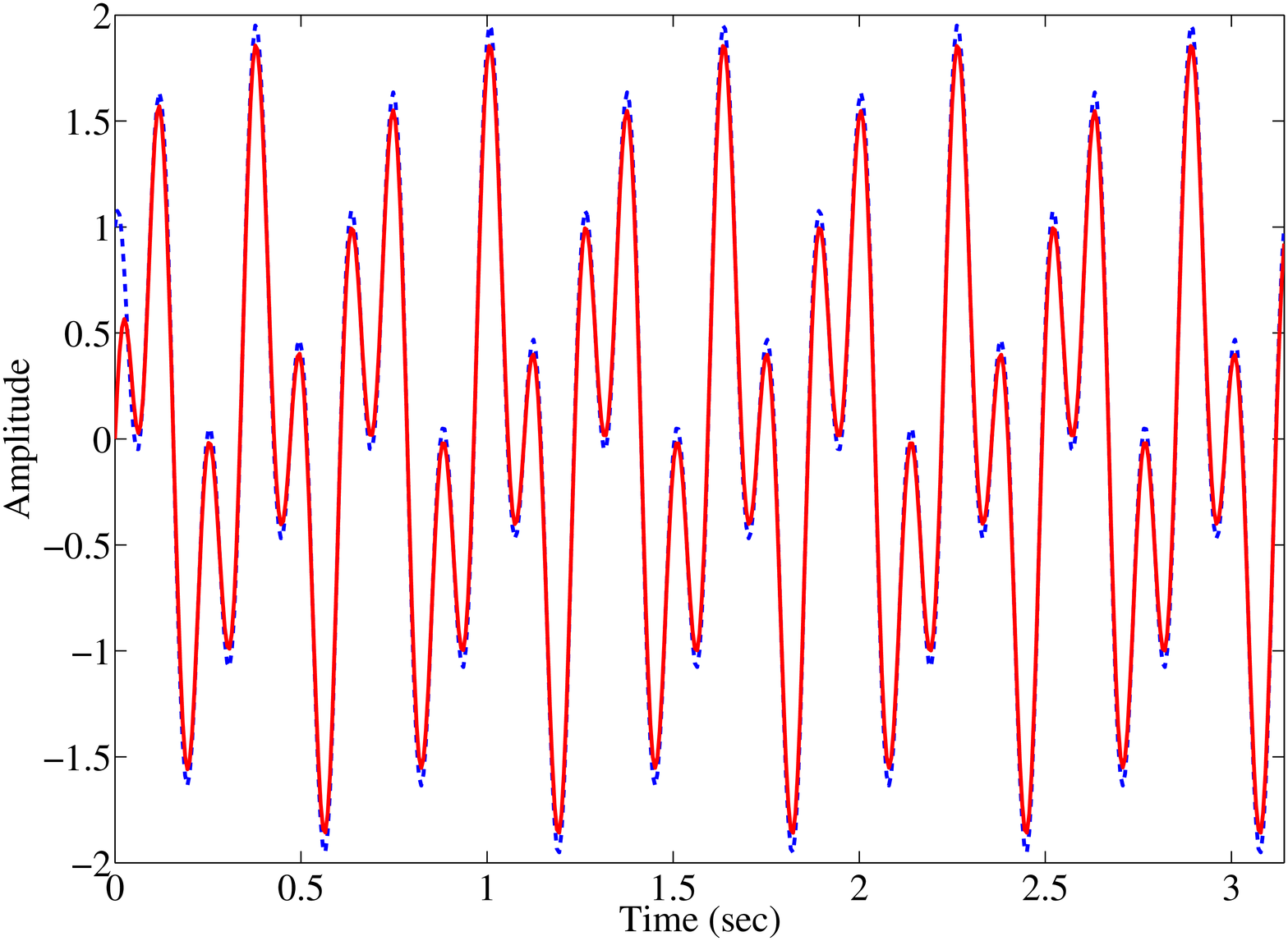}
}
\caption{
The reference $r(t)$ (dots) and the output $y(t)$ (solid), $t\in[0,\pi]$, by the $\ell^2$-optimal control}
\label{fig:output2}
\end{figure}

\begin{figure}[tb]
\centering{
\includegraphics[width=0.96\linewidth]{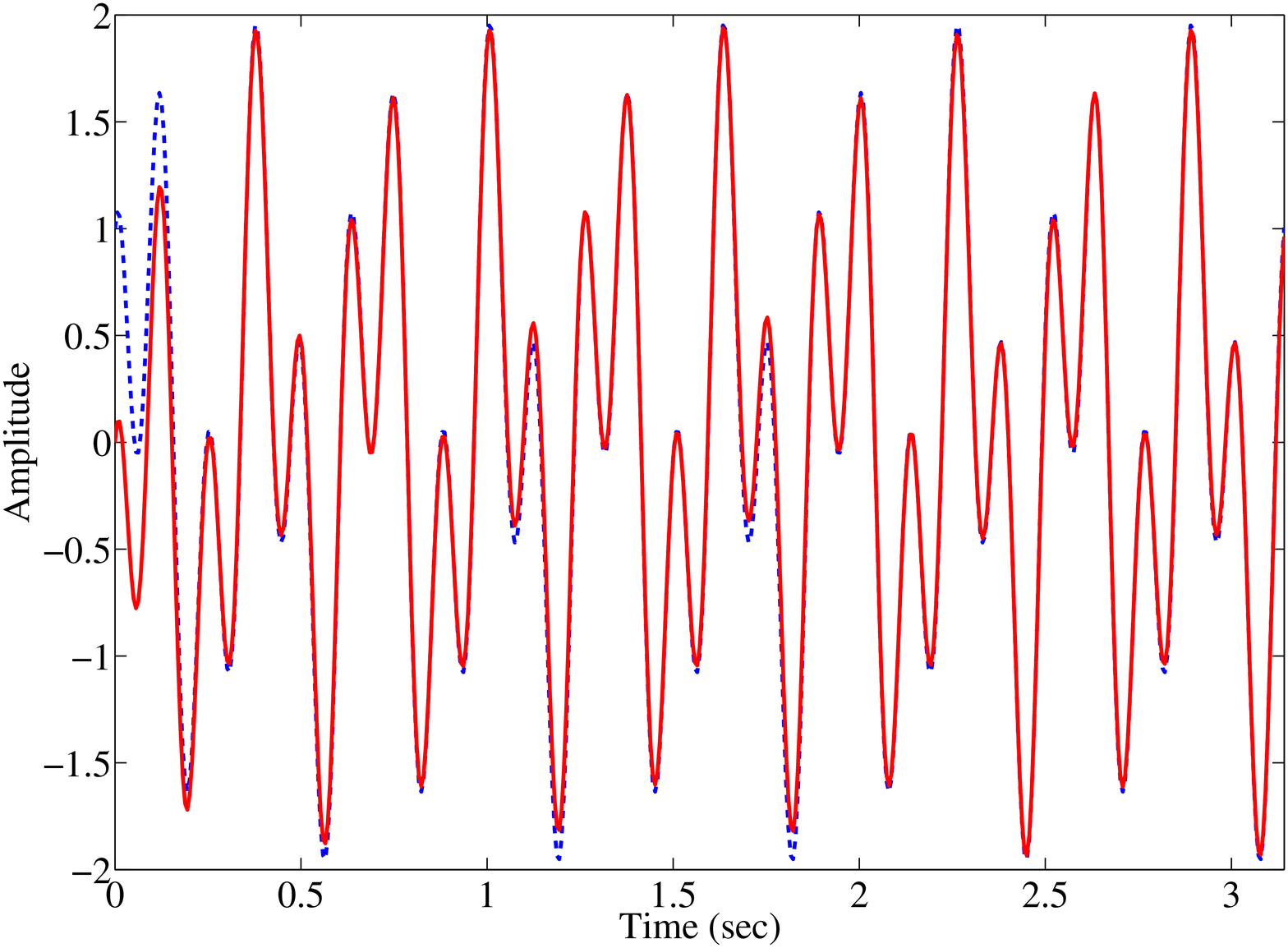}
}
\caption{
The reference $r(t)$ (dots) and the output $y(t)$ (solid), $t\in[0,\pi]$, by the $\elll$ optimal control}
\label{fig:output1}
\end{figure}

Fig.~\ref{fig:output2} shows the output $y(t)$ of the system $P$ by the $\ell^2$ optimal control.
The response is optimal in the sense that the control uses the whole sampled data on the sampling instants
$t_1,\ldots,t_{101}$.
On the other hand, Fig.~\ref{fig:output1} shows the output $y(t)$ by the proposed $\elll$ optimal control.
\begin{figure}[tb]
\centering{
\includegraphics[width=0.96\linewidth]{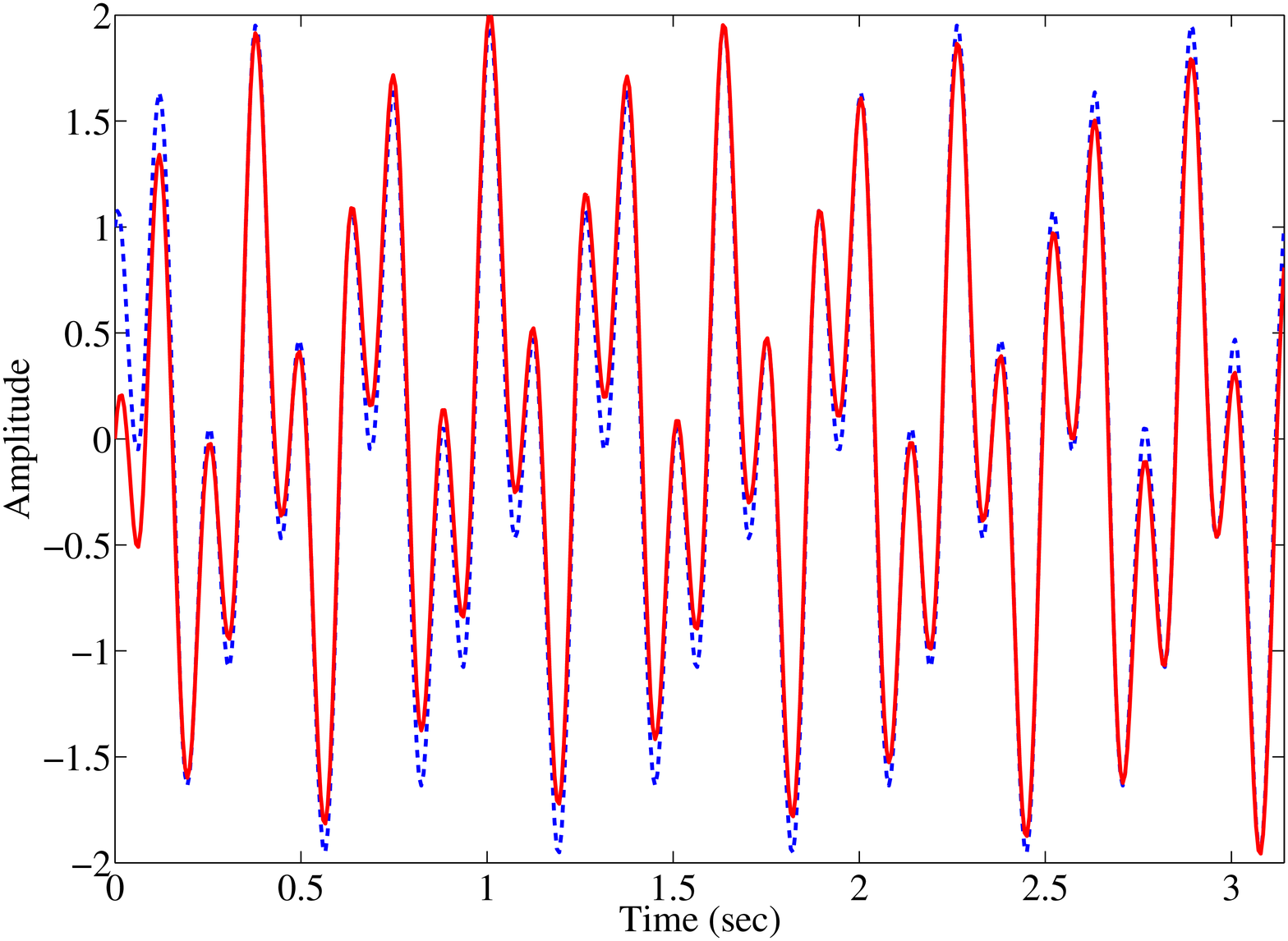}
}
\caption{
The reference $r(t)$ (dots) and the output $y(t)$ (solid), $t\in[0,\pi]$, by the truncated $\ell^2$ optimal control.}
\label{fig:outputTR}
\end{figure}
We also show the output $y(t)$ by using the $44$ largest coefficients in the $\ell^2$ optimal
vector $\ltwoopt$ (see Fig.~\ref{fig:theta2}).
Note that this truncated vector has the same cardinality as the $\elll$ optimal vector $\loneopt$.
Although the proposed control signal $\loneopt$ was computed by only $K=67$ randomly sampled data,
the output tracks the reference with quite a good performance as the $\ell^2$ optimal control,
and better than the truncation.

\begin{figure}[tb]
\centering{
\includegraphics[width=0.96\linewidth]{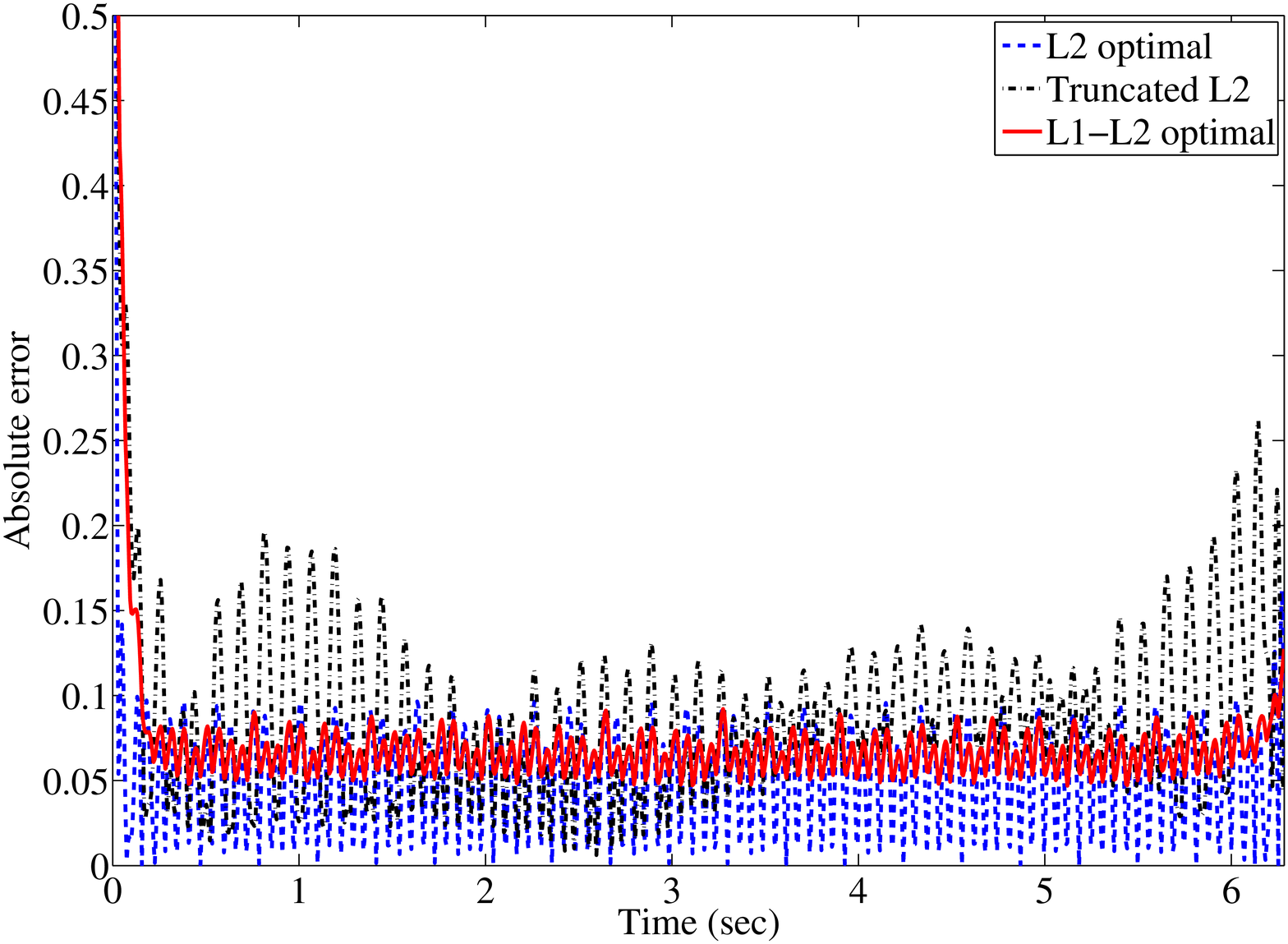}
}
\caption{The tracking error $|r(t)-y(t)|$ by the $\elll$ optimal control (averaged, solid), the $\ell^2$ optimal control(dash),
and the truncated $\ell^2$ optimal control (averaged, dash-dot) whose cardinality is the same as the $\elll$ optimal control.}
\label{fig:error}
\end{figure}
To see the difference more precisely, we run $1000$ simulations with random sampling
and compute the average of the absolute value of the tracking error $|r(t)-y(t)|$.
Fig.~\ref{fig:error} shows the result.
We can see that the control performance by the proposed method is almost comparable with that by the $\ell^2$ method,
and much better than that by the truncated $\ell^2$ optimal vector.
Note that the average of the cardinality $\|\loneopt\|_0$ is about $57.8$, 
which is about $28.8\%$ of that of $\ltwoopt$.

{\red
Then we simulate for another reference signal, the step function defined by
\[
 r(t) = 1, ~ t\in [0,2\pi].
\]
The sparsity of this reference is $S_{\!r}=1$.
We here assume that $K=N$ and run 1000 simulations with
a random initial state $\vc{x}_0\sim {\mathcal N}(\vc{0},I)$.
The other parameters are the same as above.
Fig.~\ref{fig:step_errors} shows the average of the absolute errors
by the $\ell^2$ optimal control and the $\elll$ optimal one.
\begin{figure}[tb]
\centering{
\includegraphics[width=0.96\linewidth]{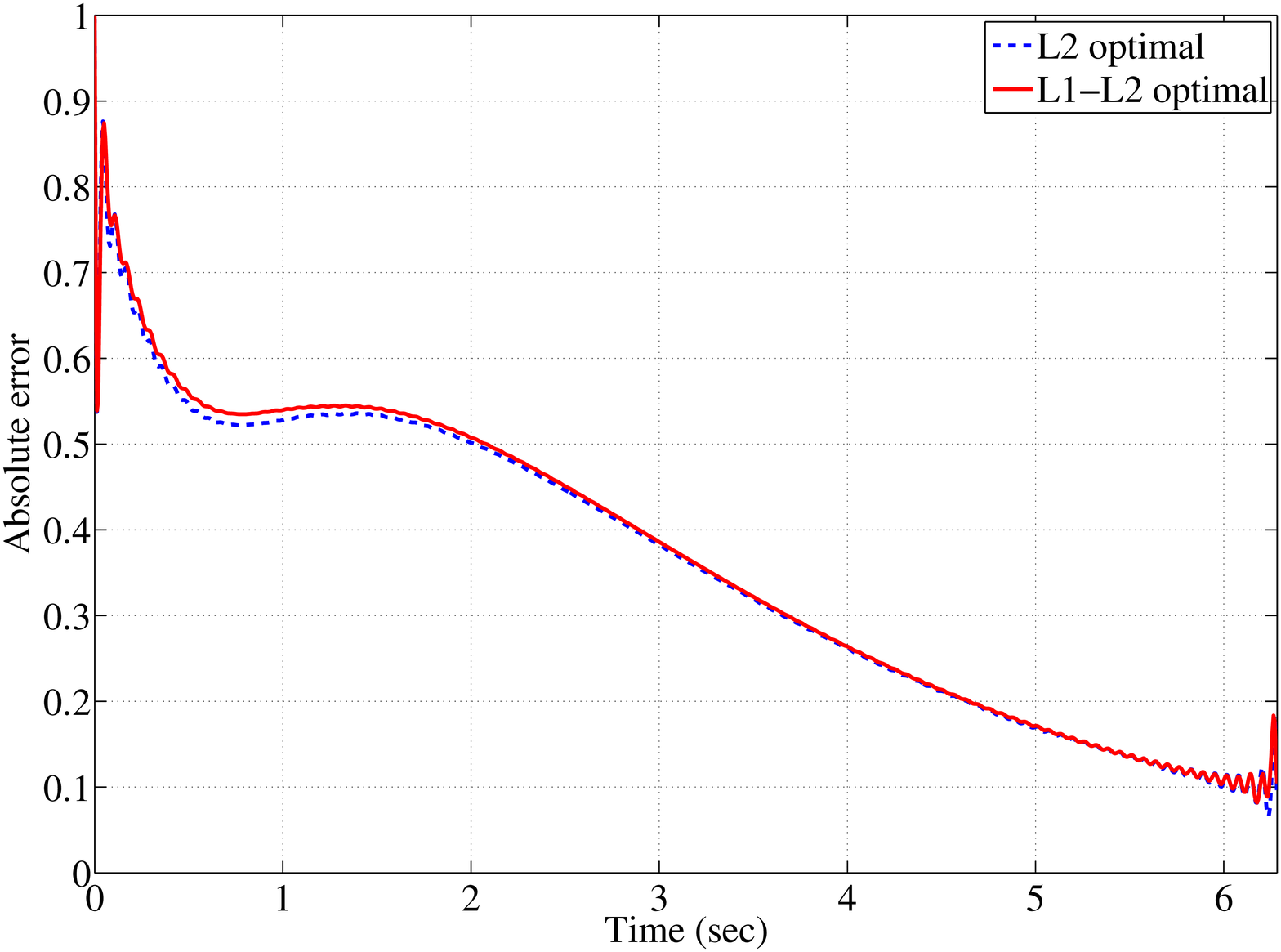}
}
\caption{\red The average of the absolute error $|y(t)-r(t)|$ of the $\ell^2$ optimal control (dash)
and the proposed $\elll$ optimal control (solid).
The performance is comparable but the proposed control vector is much sparser.}
\label{fig:step_errors}
\end{figure}
The performance is comparable but the proposed control vector $\vc{\theta}_1^\star$
has the average sparsity $\|\vc{\theta}_1^\star\|_0=152.512$, which is about $76\%$
of the full vector $\vc{\theta}_2^\star$.
That is, the proposed method can produce much sparser control vectors
without much deterioration of control performance.
}

In conclusion, the proposed method has successfully achieved an admissible level of control performance
with highly compressive sampling and sparse control signal representation.

\section{Conclusion}
\label{sec:conclusion}
In this paper, we have proposed a new method for remote control systems
based on the compressive sampling technique.
We have shown that, by assuming the sparse reference signal, 
the Fourier coefficients of the optimal tracking control signal
can be much sparser with far fewer data than what conventional design requires.
The computational cost is relatively low due to the combined use of 
the low rate random sampling and an efficient optimization algorithm.
{\red
A theoretical result has been given for control performance analysis based on the notion of 
RIP.}
Examples have been shown that the proposed method provides a very sparse control signal
without much deterioration of control performance.
{\red
The sparsity of the control vector depends also on the signal subspace $V_M$.
We leave open the problem how to select this space for a given plant $P$ and a set of reference signals.
}

\section*{Acknowledgement} This research was supported in part by
Grand-in-Aid for Young Scientists~(B) of the Ministry of Education,
Culture, Sports, Science and Technology (MEXT) under Grant No.~22760317,
No.~22700069, and No.~21760289.

\appendix

\section{\red Proof of Lemma \ref{lem:output}}
\label{sec:app:output}
{\red
For an input $u\in\V_M$, the output $y(\tau)$, $\tau\in [0,T]$ is given by
\[
 \begin{split}
  y(\tau) &= \vc{c}^\top \exp(\tau A)\vc{x}_0 + \int_0^\tau \vc{c}^\top \exp\left[(\tau-t)A\right]\vc{b} u(t) \dd t\\
  &= \vc{c}^\top \exp(\tau A)\vc{x}_0\\
  &\qquad  + \sum_{m=-M}^M \theta_m \int_0^\tau \vc{c}^\top \exp\left[{(\tau-t)A}\right]\vc{b}\psi_m(t)\dd t\\
  &= \vc{c}^\top \exp(\tau A)\vc{x}_0 + \sum_{m=-M}^M \theta_m \int_0^T \kappa(\tau,t)\psi_m(t)\dd t\\
  &= \vc{c}^\top \exp(\tau A)\vc{x}_0 + \sum_{m=-M}^M \theta_m \langle \kappa(\tau,\cdot), \psi_m \rangle.
 \end{split}
\]
}

\section{Computing inner product}
\label{sec:computing}
To compute the matrix $G$ in (\ref{eq:grammian}),
we have to compute the inner product
$\ipr{\phi_n}{\psi_m}$.
This value can be easily computed by matrix exponentials
\cite{Loa78}:
\[
 \begin{split}
  \ipr{\phi_n}{\psi_m} 
  &=\int_0^T\phi_n(t)\overline{\psi_m(t)}~\dd t\\
  &= \int_0^{t_n} \vc{c}^\top \exp\left[{(t_n-t)A}\right]\vc{b}\exp(-j\omega_m t) \dd t\\
  &= [\vc{c}^\top, 0] \exp\left(t_n\begin{bmatrix}A & \vc{b}\\ 0 & -\jj\omega_m\end{bmatrix}\right)
  \begin{bmatrix}\vc{0}_\nu\\ 1\end{bmatrix},
 \end{split}
\]
where $\vc{0}_\nu$ is the zero vector in $\comp^\nu$.
\section{FISTA}
\label{sec:app:FISTA}
We here give the algorithm of FISTA (Fast Iterative Shrinkage-Thresholding Algorithm) by \cite{BecTeb09}.

Give an initial value $\vc{\theta}[0]\in\comp^N$, and let $\beta[1]=1$, $\vc{\theta}'[1]=\vc{\theta}[0]$.
Fix a constant $c$ such that $c>\|\Phi\|^2:=\lambda_{\max}(\Phi^\top\Phi)$.
Execute the following iteration:
\[
 \begin{split}
  \vc{\theta}[j] &= \Sr_{2\mu_1/c}\left(\frac{1}{c}\Phi^\top(\vc{\alpha}-\Phi\vc{\theta}'[j])+\vc{\theta}'[j]\right),\\
  \beta[j+1] &= \frac{1+\sqrt{1+4\beta[j]^2}}{2},\\
  \vc{\theta}'[j+1] &= \vc{\theta}[j] + \frac{\beta[j]-1}{\beta[j+1]}(\vc{\theta}[j] - \vc{\theta}[j-1]),\\
  j &= 1,2,\ldots,
 \end{split}
\]
where the function $\Sr_{2\mu_1/c}$ is defined for $\vc{\theta}=[\theta_1,\ldots,\theta_N]^\top$ by
\[
  \Sr_{2\mu_1/c}(\vc{\theta}) := 
	\begin{bmatrix}
	 \sgn(\theta_1)(|\theta_1|-2\mu_1/c)_+\\
	 \vdots\\
	 \sgn(\theta_N)(|\theta_N|-2\mu_1/c)_+
	\end{bmatrix},
\]
where
$\sgn(z) := \exp(\jj \angle z)$  for $z\in\comp$,
and
$(x)_+ := \max\{x,0\}$  for $x\in\re$.

\section{\red Proof of Lemma \ref{lem:analysis}}
\label{sec:app:analysis}
{\red
Let $\vc{\theta}_1^\star(\mu_1)$ be the minimizer of the $\elll$ cost function 
$J_1(\vc{\theta})$ with the regularization parameter $\mu_1>0$.
We denote $\tilde{\vc{\theta}}^\star_1(\mu_1)$ the
reduced dimensional vector built upon the nonzero components of 
$\vc{\theta}_1^\star(\mu_1)$.
Similarly, $\tilde{\Phi}$ denotes the associated columns in the matrix $\Phi$.
By the discussion in \cite[Section IV]{Fuc04},
for sufficiently small $\mu_1$ such that 
$\mu_1\in(0,\mu_0)$, the nonempty interval in which
$\sgn(\tilde{\vc{\theta}}^\star_1(\mu_1))=\sgn(\tilde{\Phi}^+\vc{\alpha})$,
the $\elll$ optimal $\vc{\theta}^\star_1(\mu_1)$ is also the solution of
\[
 \min_\vc{\theta} \|\vc{\theta}\|_1~\text{subject to}~\|\Phi\vc{\theta}-\vc{\alpha}\|_2\leq \epsilon_2,
\]
where $\epsilon_2=\|\Phi\vc{\theta}^\star_1(\mu_1)-\vc{\alpha}\|_2$.
Then by the assumption $\delta_{2S}<\sqrt{2}-1$, we have \cite{CanRomTao06}
\[
 \begin{split}
  \|\vc{\theta}^\star_1-\vc{\theta}^\star\|_2 
   &\leq C_0 \frac{\|\vc{\theta}^\star-\vc{\theta}^\star_{[S]}\|_1}{\sqrt{S}} + C_1\epsilon_2\\
   &\leq C_0 \frac{\epsilon_1}{\sqrt{S}} + C_1\epsilon_2.
 \end{split}
\]
}

\end{document}